\documentclass[draft,showpacs,preprintnumbers,amsmath,amssymb]{revtex4}
\usepackage{graphicx}% Include figure files
\usepackage{dcolumn}% Align table columns on decimal point
\usepackage{bm}% bold math
\begin{document}
\author{\"{O}zg\"{u}r Delice}
 \email{odelice@boun.edu.tr}
 \affiliation{Department of Physics, Bo\u{g}azi\c{c}i University, 34342 Bebek, Istanbul, Turkey}
\title{Cylindrically symmetric, static
strings with a cosmological constant in Brans-Dicke theory}
\begin{abstract}
The static, cylindrically symmetric vacuum solutions with a
cosmological constant in the framework of the Brans-Dicke theory
are investigated. Some of these solutions admitting Lorentz boost
invariance along the symmetry axis correspond to local, straight
cosmic strings with a cosmological constant. Some physical
properties of such solutions are studied. These strings apply
attractive or repulsive forces on the test particles. A smooth
matching is also performed with a recently introduced interior
thick string solution with a cosmological constant.
\end{abstract}
\pacs{04.20.Jb; 04.40.Nr; 11.27.+d} \maketitle

\section{Introduction}
The Brans-Dicke (BD) theory \cite{brans-dicke} is a natural
generalization of the Einstein's general relativity (GR) where the
gravitational coupling constant is replaced by a scalar field. The
inclusion of scalar fields into GR is %also
 suggested by different
theories, for example, as a  dilaton field which naturally arises
in low energy limit of string theories or in the Kaluza-Klein
theories. This theory has some desirable properties \cite{Faraoni}
such as it obeys the weak equivalence principle and it is also
compatible with the solar system experiments for large values of
its parameter. This theory has been applied to several issues
ranging from inflation schemes in cosmology to quantum gravity
\cite{brans}.

 In this paper, we will obtain static cylindrically symmetric vacuum
solutions in the presence of a cosmological constant in the
framework of the Brans-Dicke scalar tensor theory. These solutions
can describe exterior regions of static cylindrically symmetric
sources. An important subclass of the solutions we present are the
solutions corresponding to static, straight, local cosmic strings.
Cosmic strings are linear topological defects which may have been
produced in the early universe during the phase transitions in the
general unified theories \cite{cosmicstring,Hindmarsh}. Due to
their important implications on cosmology and astrophysics, they
have attracted a lot of interest during the last decades,
especially because of the possibility that they can generate
density fluctuations in the early universe which can give rise to
the formation of galaxies, etc. Although recent observations ruled
out this possibility in favor of inflation and they cannot be the
primary source for such fluctuations, they can still contribute up
to $10\%$ \cite{wyman,bevis}, which can be detectable in the
future observations. The interest on cosmic strings was renewed
\cite{Kibble} due to several reasons, such as their relations with
fundamental strings in string theory
\cite{sarangi}-\cite{copeland} or their possible observational
consequences \cite{Sahzin}-\cite{Sazhin2} due to their
gravitational lensing effects. In the framework of scalar-tensor
theories, cosmic strings are investigated in detail
\cite{Pimentel}-\cite{Lee}.

Strings with a cosmological constant in GR theory  is presented by
Linet \cite{Linet} and Tian \cite{tian}. Note that, the solutions
with a cosmological constant for a diagonal metric which  depends
only on one variable were studied for GR \cite{SKMHH} and BD
\cite{Petzold} theories before. The solutions we will discuss, in
certain limits, reduce to exterior cosmic strings with
\cite{Linet,tian} or without \cite{cosmicstring} cosmological
constant in GR, or cosmic strings in BD theory
\cite{GundlachOrtiz}-\cite{BoisseauLinet}. Moreover, some
particular solutions we will discuss do not have a GR analogue.
Since there are evidences that we can have a nonzero cosmological
constant in the present universe \cite{Riess}, and different
theories \cite{brans} suggest that we can have a scalar field
nonminimally coupled to gravity, it might be interesting to find a
solution describing cosmic strings in the presence of cosmological
constant in a scalar tensor theory of gravitation.

The paper is organized as follows. In Sec. \ref{Sec2} and
\ref{Sec3}, we will introduce the field equations for a general
static cylindrically symmetric ansatz and their different general
solutions. Then, in Sec. \ref{Sec4}, we will select the solutions
corresponding to static, straight cosmic strings among these
solutions. Some of the physical properties of such solutions will
be discussed in Sec. \ref{Sec5} and \ref{Sec6}. In Sec.
\ref{Sec7}, by matching a recently introduced regular interior
thick string solution \cite{Delice} smoothly, we will show that
these exterior string solutions can be generated by cosmic
strings. Then, we finish with a brief discussion.

\section{Field Equations\label{Sec2}}
In the Jordan-Fierz frame, the Brans-Dicke theory is described, in
the presence of a cosmological constant, by the action:
\begin{eqnarray}\label{BDaction}
S&=&\int d^4x
\sqrt{-g}\left\{\phi(-R+2\Lambda)+\frac{\omega}{\phi}g^{\mu\nu}\partial_\mu\phi
\partial_\nu \phi\right\}%\nonumber\\&&
+  S_m[\Psi,g],
\end{eqnarray}
here $R$ is the Ricci scalar, $\Lambda$ is cosmological constant,
$g$ is determinant of the spacetime metric $g_{\mu\nu}$, $\phi$ is
the Brans-Dicke scalar field, $\omega$ is Brans-Dicke parameter,
and $S_m$ denote action of matter fields $\Psi.$ We use units in
which $c=\hbar=1$ and $(-+++)$ signature. This theory is metric,
since the matter fields are universally coupled to gravity. The
action (\ref{BDaction}) yields the field equations as:
\begin{eqnarray}
 \label{metric-eqn} G_{\mu\nu}+\Lambda
g_{\mu\nu}&=&\frac{T_{\mu\nu}}{\phi}+\frac{\omega}{\phi^2}\left(
\phi_{,\mu}
\phi_{,\nu}-\frac{1}{2}g_{\mu\nu}\phi^{,\alpha}\phi_{,\alpha}
\right)%\nonumber
%\\&&
+\frac{1}{\phi}\left(\phi_{,\mu;\nu}-g_{\mu\nu}\Box\phi
\right),\phantom{aa}\\
\label{scalar-eqn}(2\omega+3)\Box\phi&=&-2\Lambda \phi+T.
\end{eqnarray}
Here $T_{\mu\nu}$ is the energy-momentum tensor of the matter
fields and $T=T^{\mu}_{\phantom{a}\mu}$ is its trace.  In this
frame (\ref{BDaction}), the energy conservation equation
\begin{equation}\label{encons}
\nabla_\mu T^\mu_{\phantom{a}\nu}=0,
\end{equation}
 holds. Note that it is also possible to formulate the Brans-Dicke
 theory in a frame (Einstein frame) where scalar field is only coupled matter
 fields. However in this paper we will only consider Jordan-Fierz % Brans-Dicke
 frame (\ref{BDaction})  where test particles
 follow geodesics and energy conservation equation holds.

We now consider a cylindrically symmetric static metric
\begin{equation}
ds^2=-g_{0}^2dt^2+dr^2+g_{2}^2dz^2+g_{3}^2d\varphi^2,
\end{equation}
where the metric functions $g_i, (i=0,2,3)$ and scalar field
$\phi$ depend on the coordinate $r$. The coordinates
$t,r,z,\varphi$ are the time-like, the radial, the axial and the
angular coordinates of the cylindrical symmetric static metric
with ranges $-\infty<t,z,<\infty, 0\le r \le \infty$ and $0 <
\varphi \le 2\pi$ respectively.

Using the metric ansatz given above and taking $U=g_0g_2g_3$ and
$W=U\phi$, after some arrangements, the vacuum ($T_{\mu\nu}=0$) BD
field equations can be written as:
\begin{eqnarray}
&&\left(\frac{g_i'}{g_i}\,W\right)'=
-2\,\Lambda\,\frac{\omega+1}{2\omega+3}W,\label{EqA}\\
&&\left(\frac{\phi'}{\phi}\,W\right)'=-\frac{2\Lambda}{2\omega+3}\,W,\label{EqP}\\
&&\frac{g_0'g_2'}{g_0g_2}+\frac{g_0'g_3'}{g_0g_3}+\frac{g_2'g_3'}{g_2g_3}+\frac{\phi'U'}{\phi
U} =\omega\frac{\phi'^2}{2\phi^2}-\Lambda.\label{EqGrr}
\end{eqnarray}
From (\ref{EqA}) and (\ref{EqP})  we find:
\begin{equation}\label{EqW}
W''+\alpha W=0,\
\end{equation}
where
\begin{equation}\label{alpha}
\alpha=2\Lambda\,\frac{3\omega+4}{2\omega+3}.
\end{equation}
Note that $W$ satisfies
\begin{equation}
W'^2=-\alpha W^2+k^2,\quad k=\mbox{constant.}
\end{equation}
A similar analysis was presented in \cite{Petzold} for a diagonal
Bianchi Type I universe.

\section{Solutions of the field equations\label{Sec3}}
Here, we will investigate the solutions of the equations
(\ref{EqA})-(\ref{EqGrr}) for the cases $\alpha=0$, $\alpha>0$,
and $\alpha<0$, separately.

\subsection{Solutions for $\alpha \neq 0$, $\omega\neq -3/2$ }
The equations (\ref{EqA}-\ref{EqP}) can be integrated to give (for
$\alpha \neq 0$):
\begin{eqnarray}\label{EqAA}
\frac{g_i'}{g_i}=\frac{(\omega+1)}{(3\omega+4)}\frac{W'}{W}+\frac{k
c_i}{W}\\
\frac{\phi'}{\phi}=\frac{1}{(3\omega+4)}\frac{W'}{W}+\frac{k
c_4}{W}\label{EqPP}
\end{eqnarray}
where $c_i,c_4$ are constants. Putting these into the
(\ref{EqGrr}) one finds the constraints that these constants must
satisfy as:
\begin{eqnarray}
&&c_0+c_2+c_3+c_4=0,\label{constr1}\\
&&c_0^2+c_2^2+c_3^2+(1+\omega)c_4^2=\frac{2\omega+3}{3\omega+4}\label{constr2}.
\end{eqnarray}

Now, let us introduce the metric functions. The equation
(\ref{EqW}) has solutions $(\omega\neq -3/2)$
\begin{eqnarray}
&&W(r)=\gamma \sin(\sqrt{\alpha} r)+\delta \cos(\sqrt{\alpha}
r),\quad
(\alpha>0),\phantom{aaa} \label{W1}\\
&&W(r)=\gamma \sinh(\sqrt{|\alpha|} r)+\delta
\cosh(\sqrt{|\alpha|} r),\ (\alpha<0).\phantom{aaa} \label{W2}
\end{eqnarray}
Cylindrical symmetry imposes that we have to choose $\delta=0$. %in
Using these, the equations (\ref{EqAA}-\ref{EqPP}) can be easily
solved.
\subsubsection{$\alpha>0$\label{SolnLg0}} For this case we find that:
\begin{eqnarray}
&&g_i(r)=C_i\tan{(\sqrt{\alpha}r/2)}^{c_i}\sin{(\sqrt{\alpha}r)}^{\frac{\omega+1}{3\omega+4}},\\
&&\phi(r)=\phi_0\tan{(\sqrt{\alpha}r/2)}^{c_4}\sin{(\sqrt{\alpha}r)}^{\frac{1}{3\omega+4}}.
\end{eqnarray}

\subsubsection{$\alpha<0$\label{SolnLs0}}
for this case we have
\begin{eqnarray}\label{metaneg}
&&g_i(r)=C_i\tanh{(\sqrt{|\alpha|}r/2)}^{c_i}\sinh{(\sqrt{|\alpha|}r)}^{\frac{\omega+1}{3\omega+4}},\phantom{aa}\\
&&\phi(r)=\phi_0\tanh{({|\alpha|}r/2)}^{c_4}\sinh{(\sqrt{|\alpha|}r)}^{\frac{1}{3\omega+4}}.\phantom{a}\label{metpneg}
\end{eqnarray}
For both of the solutions, the constants $c_i$ must satisfy
(\ref{constr1},\ref{constr2}) and also $C_0C_2C_3\phi_0=\gamma$.
Actually, we can take $C_0=C_2=1$ by rescaling the coordinates $t$
and $z$. Note that for both of the solutions, the limits
$c_4\rightarrow 0,$ $w\rightarrow \infty$,
$c_4/\sqrt{\omega}\rightarrow 0$, yield their GR counterparts
\cite{Linet}.

\subsection{Solutions for $\alpha=0$, $\omega\neq -3/2$}
For $\alpha=0$ we have $W''=0$ and we can take $W=r$ where $r=0$
is chosen as the symmetry axis. We have two different solutions
for  the special cases $\omega=-4/3$ and $\Lambda\neq0$ or
$\Lambda=0$ and $\omega$ arbitrary.
\subsubsection{A special solution for $\omega=-4/3$, $\Lambda\neq
0$\label{soln-43}} By considering (\ref{EqA}) and (\ref{EqP}) we
find:
\begin{eqnarray}
&&g_i(r)=B_i r^{b_i} e^{\Lambda r^2/2},\\
&&\phi=B_4 r^{b_4} e^{-3\Lambda r^2/2}.
\end{eqnarray}
Note that the constants $b_i,b_4$ and $B_i,B_4$  are not
arbitrary. By considering the definition of $W$ we have
$B_0B_2B_3B_4=\gamma$ and we can take $B_0=B_2=1$ by rescaling the
coordinates $t$ and $z$. From the equation (\ref{EqGrr}) we find
that these constants should satisfy the following conditions:
\begin{eqnarray}
&&b_0+b_2+b_3+b_4=1,\\
&&b_0^2+b_2^2+b_3^2-b_4^2/3=1.
\end{eqnarray}
Note that this special solution has no general relativistic limit
for $\Lambda\neq 0$.
\subsubsection{General solution with $\Lambda=0$, $\omega\neq -3/2$}
For this case the solution reduces to the cylindrical vacuum
solution in Kasner form with
\begin{eqnarray}
g_i(r)=A_i r^{a_i},\phi=A_4 r^{a_4},
\end{eqnarray}
where the constants $A_i,A_4$ have to satisfy
$A_0A_2A_3A_4=\gamma$. Here we can take $A_0,A_2$ as unity and the
constants $a_i,a_4$ have to satisfy the Kasner conditions for
Brans-Dicke theory:
\begin{eqnarray}
a_0+a_2+a_3+a_4=1,\\
a_0^2+a_2^2+a_3^2+(1+w)a_4^2=1.
\end{eqnarray}
The limit $a_4\rightarrow 0,$ $\omega\rightarrow \infty$,
 $a_4/\sqrt{\omega} \rightarrow 0$ leads to general relativistic version of the
solution, namely the Levi-Civita solution \cite{levicivita} in
Kasner form.

\section{Solutions corresponding to Cosmic Strings\label{Sec4}}
By selecting the solutions admitting Lorentz boosts in $t,z$
directions, we can identify the solutions corresponding to
strings.
\subsection{For $\alpha\neq 0$}
For this case we need $c_0=c_2$ and the constraint equations
reduce to the following conditions:
\begin{eqnarray}
&&c_3=-2c_0-c_4,\label{const1}\\
&&c_0=-\frac{1}{6}\left(2c_4\pm\sqrt{2}\sqrt{\frac{3(2\omega+3)}{3\omega+4}-c_4^2}
\right)\label{const2}.
\end{eqnarray}
Applying these relations, for $\Lambda<0$, the metric becomes
\begin{eqnarray}
ds^2&=&dr^2+A(r)^2B(r)^{2c_0}\bigg(-dt^2+dz^2%\nonumber \\
%&&\phantom{aaaaaaaaaaaaa}
+C_3^2B(r)^{\pm 2c_5}d\varphi^2\bigg),
\label{metriccompact}
\end{eqnarray}
with
\begin{eqnarray}
A(r)&=&\sinh(\sqrt{|\alpha|}r)^{\frac{\omega+1}{3\omega+4}},\\
B(r)&=&\tanh(\sqrt{|\alpha|}r/2).\phantom{aa}\\
c_5&=&\frac{1}{\sqrt{2}}\sqrt{\frac{3(2\omega+3)}{3\omega+4}-c_4^2},
\end{eqnarray}
and the scalar field is given in (\ref{metpneg}). For $\Lambda>0$,
hyperbolic functions in the functions above should be replaced by
corresponding trigonometric functions. For cosmic strings in BD
with $\Lambda \neq 0$, the solution has four parameters:
$\Lambda,\omega,c_4,C_3$. Since metric must be real, for
$-3/2<\omega<-4/3$ there is no solution and $c_4$ must satisfy the
relation $0\le c_4^2 \le 3(2\omega+3)/(3\omega+4)$.
 These solutions reduce to  strings with cosmological constant
in GR limit, since in that limit ($c_4\rightarrow 0$, $\omega
\rightarrow \infty$) we have $(c_0,c_2,c_3)=(\pm 1/3,\pm 1/3,\mp
2/3)$.
\subsection{For $\alpha=0$}
\subsubsection{$\omega=-4/3$, $\Lambda\neq 0$}
For this case we have $b_0=b_2$ and the constraint equations
reduce to:
\begin{eqnarray}
b_3=1-2b_0-b_4\\
b_0=\frac{1}{3}\left(1-b_4\pm\sqrt{1+b_4}\right).
\end{eqnarray}
Thus we must have $b_4>-1$. These solutions have no GR limit for
$\Lambda\neq 0$ and reduce to the exterior string solutions for
$\Lambda=b_4=0.$
\subsubsection{$\Lambda=0$} For this case we have $a_0=a_2$ and the
constraint equations take the form:
\begin{eqnarray}
&&a_3=1-2a_1-a_4,\\
&&a_1=\frac{1}{3}\left(1-a_4\pm\sqrt{1+a_4-\frac{a_4^2}{2}(3\omega+4)}
\right)
\end{eqnarray}
and relations $1-\sqrt{3(2\omega+3)}/(3\omega+4)\le a_4 \le
1+\sqrt{3(2\omega+3)}/(3\omega+4)$ and $\omega>-3/2$ must be
satisfied. These are equivalent to the well known exterior cosmic
string solutions
\cite{GundlachOrtiz,BoisseauLinet},\cite{DahiaRomero} for the BD
theory. Note that a conformal transformation is needed since those
solutions are presented in the conformal or Einstein frame, which
is related with the BD frame by a conformal transformation.
 For the GR limit, these conditions give us two different solutions. The first one is the famous cosmic string
 solution of Vilenkin $(a_0,a_2,a_3)=(0,0,1)$ and the second one is  related with supermassive or global strings
 $(a_0,a_2,a_3)=(2/3,2/3,-1/3)$.

\section{Singularity Behaviour\label{Sec5}}

Let us first consider the general solution
(\ref{const1}-\ref{const2}). First of all, GR strings do not have
curvature singularity on the axis \cite{Linet}, whereas BD strings
have, due to the interaction of BD scalar. Secondly, unlike
$\Lambda<0$ case, GR strings with $\Lambda>0$ are singular at
$r=r_g=\pi/\sqrt{3\Lambda}.$ For the BD theory we have the similar
case and for $\alpha>0$ the solutions are singular at
$r=r_b=\pi/\sqrt{\alpha}$ where $\alpha$ is given in
(\ref{alpha}). Actually, for BD case the solutions depend on the
sign of $\alpha$ rather than $\Lambda$. However, since the cases
that they have different signs ($-3/2<\omega<-4/3$) are excluded,
otherwise the metric functions become imaginary for this interval,
we cannot have solutions which $\alpha$ and $\Lambda$ have
different signs. Thus, for GR and BD theories, the solutions with
$\Lambda>0$ are singular at $r=r_b$. Due to singularity at $r=r_b$
either the solutions with $\Lambda>0$ must be excluded or
following Tian \cite{tian}, they can describe a string in a closed
universe whose topology is $R^2\times S^2$. This can be achieved
by transforming the radial coordinate as $r=\chi+\chi_0$ where
$\chi_0=r_b=\pi/\sqrt{\alpha}$ and the coordinate $\chi$ have the
range $\chi_0\le\chi\le\pi/\sqrt{\alpha}$.

Another important difference is the special solution
(\ref{soln-43}) which has no GR analogue for $\Lambda\neq 0$. This
solution is also singular on the axis in general and  it is not
singular for finite $r$ unless $r\rightarrow \infty$. We can avoid
the singularity at the radial infinity by applying the
prescription given in the previous paragraph. For the following
choice of parameters, we can have a metric similar to cosmic
string metric in GR for $b_4=b_0=b_1=0,b_3=1$:
\begin{eqnarray}
ds^2&=&dr^2+e^{\Lambda r^2}\left(-dt^2+dz^2+B_3^2\,r^2d\varphi^2
\right),\\
\phi&=&B_3^{-1}e^{-3\Lambda r^2/2}.
\end{eqnarray}
This solution is not singular at $r=0$, but it is singular at
$r\rightarrow \infty$. Its Ricci scalar is
$R=-2(7\Lambda+6\Lambda^2r^2)$ and the other curvature scalars
have the same behavior.
\section{Motion of test particles\label{Sec6}}

Hereafter we only consider the solution with $\Lambda>0$, which is
not singular anywhere except on the axis. We investigate geodesic
equations for a test particle moving at $z=\mbox{constant}$ planes
for  the metric (\ref{metriccompact}).
The geodesic equations %for $z=\mbox{constant}$ planes
 are:
\begin{eqnarray}
\ddot{r}&=&-A^2B^{2c_0}\bigg\{\left(\frac{A'}{A}+c_0\frac{B'}{B}\right)\dot{t}^2
%\nonumber\\
%&&\phantom{AAA}
-C_3^2B^{\pm 2c_5}\left(\frac{A'}{A}+(c_0\pm c_5)
\frac{B'}{B}\right)\dot{\varphi}^2\bigg\}
\\
\dot{t}&=&\frac{t_0}{ A^2B^{2 c_0}},\label{geodesict}\\
\dot{\varphi}&=&\frac{\varphi_0}{A^2B^{2(c_0-c_5)}},\label{geodesicp}
\end{eqnarray}
where $t_0$ and $\phi_0$ are integration constants and overdot
represents differentiation with respect to proper time $\tau.$.
From the metric, we have,
\begin{equation}\label{geodesicmet}
\dot{r}^2+A^2B^{2c_0}\left(-\dot{t}^2+C_3^2B^{\pm2c_5}
\dot{\varphi}^2\right)=\epsilon,
\end{equation}
where $\epsilon=-1$ for timelike particles and $\epsilon=0$ for
photons. The equation (\ref{geodesicmet}), with the help of
equations (\ref{geodesict}) and (\ref{geodesicp}) gives
\begin{eqnarray}
\left(\frac{dr}{d\varphi}\right)^2&=&\frac{\dot{r}^2}{\dot{\varphi}^2}
=A^2B^{2c_0-4c_5}%\times
%\nonumber\\
%&&\phantom{a}
\bigg[\left(\frac{t_0}{\varphi_0}\right)^2+C_3^2B^{\pm2c_5}\left\{
\frac{\epsilon}{\varphi_0^2}-1 \right\} \bigg]\label{radgeo}
\end{eqnarray}
The expression inside the square brackets must be positive since
left hand side of the equality is a perfect square.

Let us consider photons ($\epsilon=0$). When we choose positive
sign in (\ref{radgeo}) then we have an increasing function of $r$,
but since this function is limited from above, depending of the
value of $t_0/\varphi_0$, the photons emitted from surface of a
cylinder can reach to radial infinity or not. Moreover, the
photons travelling from infinity towards the string can reach it.
However for the minus sign in (\ref{radgeo}) we have a decreasing
function of $r$ and photons emitted from surface of a cylinder can
reach the radial infinity. However, the photons travelling from
infinity towards the string cannot reach it since there exist an
$r=r_0$ where the term inside the parenthesis becomes negative.
Thus, for this case we have a barrier at $r=r_0$, hence string has
repulsive effect for this case.

For timelike particles we consider radial acceleration of a test
particle at rest and it is given by:
\begin{eqnarray}
\ddot{r}&\approx&-\left(\frac{A'}{A}+
c_0\frac{B'}{B}\right)\nonumber\\
&\approx&
-%\frac{
\left(c_0+\frac{(\omega+1)}{(3\omega+4)}\cosh{(\sqrt{|\alpha|}r)}
\right)%}{
\sinh{(\sqrt{|\alpha|}r)}.\phantom{aa}
%}
\label{racc}.
\end{eqnarray}
When $\ddot{r}$ is negative (positive) the string applies
attractive (repulsive) force  whereas when it vanishes it applies
no force on a test particle initially at rest. We see that the
behaviour of the force depends on the terms inside the
parenthesis: the constant $c_0$ and the term
\begin{equation}
\kappa(r)=(\omega+1)/(3\omega+4)\cosh{(\sqrt{|\alpha|}r)}.\label{kappa}
\end{equation}
When $c_0$ positive, if $\kappa(r)$ is also positive ($\omega>-1$
or $\omega<-4/3$) string applies attractive force everywhere,
whereas if $\kappa(r)$ is negative ($\omega \in (-1,-4/3)$) , the
force becomes repulsive whenever $|\kappa|>c_0$. For negative
$c_0$ the situation is reverse and if $\kappa(r)$ is also negative
the force is repulsive everywhere, whereas for positive $\kappa$
the force becomes attractive for larger $r$ when $|c_0|<\kappa$.

\section{Interior solutions\label{Sec7}}
The interior region of a static straight local cosmic string can
be modelled by an anisotropic fluid having the energy momentum
tensor of the form:
\begin{equation}
T^\mu_{\phantom{a}\nu}=\mbox{diag}(-\rho,p_r,p_z,p_\varphi),
\end{equation}
Here $\rho$ is the energy density and $p_r$, $p_z$, and
$p_\varphi$ are the principal pressures. In GR theory with no
cosmological constant, $\Lambda=0$, local cosmic strings can be
characterized by a phenomenological equation of state  due to
Vilenkin \cite{Vilenkin}:
\begin{eqnarray}
\rho=-p_z,\quad p_r=p_\varphi=0.
\end{eqnarray}
 However, in the presence of a cosmological constant in GR
theory, we have to take into account other pressures
($p_r,p_\phi$) as nonvanishing \cite{tian}, otherwise we do not
have any solution apart from vacuum one. This is also the case for
the Brans-Dicke theory without $\Lambda$ , since Sen et. al.
showed \cite{Sen} that the Vilenkin's ansatz $\rho=-p_z$ , with
other pressures are vanishing is inconsistent. This inconsistency
can be removed for different cases such as for nonstatic strings
\cite{Senns},  or for a more general scalar tensor theory
\cite{SenBanerjee}. Furthermore,  in a recent paper \cite{Delice},
it has been shown that,  in the presence of a cosmological
constant, this ansatz for interior region is indeed consistent.
Some interior solutions having above energy momentum tensor is
also presented in that work. One of the solutions for $\Lambda<0$
is given by:
\begin{eqnarray}
ds^2&=&-dt^2+dr^2+dz^2+W^2d\varphi^2,\label{interior}\\
W&=&C \sinh((\sqrt{|\beta|}/2)\,r)
\cosh^{\frac{\omega}{\omega+2}}((\sqrt{|\beta|}/2)\,r),\\
\phi&=&\frac{A}{C} \cosh^{\frac{2}{\omega+2}}((|\beta|/2)\,r),\\
\beta&=&2\Lambda(\omega+2).
\end{eqnarray}
For this solution, the energy density is given by
\begin{equation}
\rho=-p_z=2(\omega+1)\phi\Lambda.
\end{equation}
This solution has some remarkable features such as it is smooth
and regular on the axis and it has no curvature singularities. The
GR limit of this solution is given by $\omega\rightarrow \infty$,
$\Lambda\rightarrow 0$ and $\Lambda\omega\rightarrow
\mbox{constant}$. Hence it has no GR limit for $\Lambda\neq 0.$

The mass per unit length of the string is given by
\begin{eqnarray}
\mu=2^{(2\omega+2)/{(\omega+2)}}\pi\frac{(\omega+1)}{(\omega+2)}\sinh^2{\left((\beta/2)\,r_0\right)}.
\end{eqnarray}

For a more consistent treatment, we have to match this interior
solution with exterior solutions (\ref{metriccompact}) and
(\ref{metpneg}) smoothly. We match these metrics at a hypersurface
defined by $\Sigma: r_{\pm}=r_{0\pm}$. For a smooth matching, we
need \cite{Barrabes}:
\begin{eqnarray}
&& [g_{ij}]=0,\quad [K_{ij}]=0,
 \quad {[\phi ]}=0,\quad {[ \phi'
]}=0,
\end{eqnarray}
where $[X]=X_+-X_-$ is a jump of a quantity on the hypersurface
$\Sigma$ and $K_{ij}$ is the extrinsic curvature tensor. We choose
same $t,z,\varphi$ coordinates  for the interior (\ref{interior})
and the exterior solutions (\ref{metriccompact}) except for the
radial coordinates.  The interior (exterior) radial coordinate is
denoted by $r_-$ ($r_+$). The junction conditions give us the
following equations:
\begin{widetext}
\begin{eqnarray}
&&C_0=\tanh^{-c_0}{\left(\sqrt{|\alpha|}r_{0+}/2\right)} \sinh^{-\frac{\omega+1}{3\omega+4}}\left({\sqrt{|\alpha|}r_{0+}}\right),\\
&&C_3=C\tanh^{-(c_0\pm c_5)}{\left(\sqrt{|\alpha|}r_{0+}/2\right)} \sinh^{-\frac{\omega+1}{3\omega+4}}\left({\sqrt{|\alpha|}r_{0+}}\right) %\nonumber \\
 \sinh{\left(\sqrt{|\beta|} r_{0-}/2 \right)}\cosh^{\frac{\omega}{\omega+2}}{\left(\sqrt{|\beta|} r_{0-}/2\right)},\\
&&\phi_0=\frac{A}{C}\tanh^{-c_4}{\left(\sqrt{|\alpha|}r_{0+}/2\right)} \sinh^{-\frac{1}{3\omega+4}}\left({\sqrt{|\alpha|}r_{0+}}\right)%\nonumber \\
\cosh^{\frac{2}{\omega+2}}(\sqrt{|\beta|}\,r_{0-}/2), \\
&&c_0=-\frac{1+\omega}{3\omega+4}\cosh{(\sqrt{|\alpha|}\, r_{0+})},\label{junctc0}\\
&&\frac{\sqrt{|\alpha|}}{C\sqrt{|\beta|}}\left(c_0\pm c_5 +
\frac{\omega+1}{3\omega+4}\cosh{(\sqrt{|\alpha|}r_{0+})}\right)
\frac{(\omega+2)\,\sinh(\sqrt{|\beta|}r_{0-})}{\cosh^{\frac{-2}{\omega+2}}{(\sqrt{|\beta|}r_{0-}/2)}\sinh{(\sqrt{|\alpha|}r_{0+})}}%\nonumber\\
%&&\phantom{aa}
=
1+(\omega+1)\cosh{(\sqrt{|\beta|}r_{0-})},\phantom{aaa} %\right)
\\
&&\sqrt{|\alpha|}\left(c_4+\frac{\cosh{(\sqrt{|\alpha|}r_{0+}})}{3\omega+4}\right)
\coth({\sqrt{|\beta|}r_{0-}/2})
=\frac{\sqrt{|\beta|}}{\omega+2}\sinh{(\sqrt{|\alpha|}r_{0+})}.
\end{eqnarray}

\end{widetext}
Since $c_0$ and $c_5$ in (\ref{metriccompact}) depends on $c_4$
and $\omega$ (\ref{const2}), we have six equations for nine
unknowns $C_0,C_3,\phi_0,C,c_4,\alpha,\beta,r_{0-},r_{0+}$. Thus,
from the junction conditions, the exterior metric and scalar field
parameters can be determined in terms of the interior metric and
scalar field  parameters $r_{0-},\beta,C$. Hence, these junction
conditions are fulfilled. This yields a configuration with a thick
string on the axis with vacuum exterior with $\Lambda$ in BD
theory.

\section{Discussion\label{Sec8}}
In this paper cylindrically symmetric static solutions of the
Brans-Dicke field equations in the presence of a cosmological
constant are investigated. Among these, the solutions admitting
boost invariance along the symmetry ($z$) axis can represent the
exterior fields of cosmic strings with $\Lambda$. Some physical
properties of these solutions such as their singularity behaviour
and motion of test particles are discussed. Then, we have shown
that it is possible to match these exterior solutions with an
regular interior string solution with $\Lambda$ \cite{Delice},
having a phenomenological equation of state $\rho=-p_z$, smoothly.

In BD theory with a cosmological constant, as in GR theory with
$\Lambda$, local strings have nonvanishing gravitational potential
and affects the motion of test particles. Our investigations in
Sec. \ref{Sec6} show that, these strings can apply either an
attractive or repulsive force on a test particle initially at
rest, depending of the parameters $c_0,|\Lambda|,\omega$ and the
radial distance from the string. The junction condition
(\ref{junctc0}) yields that we have $c_0=-\kappa(r_{0+})$ where
$\kappa(r)$ is given in Eq. (\ref{kappa}) and $r_{0+}$ is the
thickness of the string according to exterior metric
(\ref{metriccompact}). Hence, $c_0$ is negative for all values of
$\omega$ except for $\omega\in (-4/3,-1)$ and the gravitational
behaviour of the string is directly related to its thickness and
its tension.

Strings with negative cosmological constant are well behaving
everywhere, they are only singular at the origin, but since there
is an interior solution occupying the axis and smoothly matching
to the exterior, this is not a problem. However, for the strings
with positive cosmological constant, there is a singularity from
the finite distance from the axis. Hence, a string with  a
positive cosmological constant cannot be physically acceptable
unless the space-time has a closed topology.

\end{document}